\documentclass{emulateapj}
\usepackage{booktabs}
\usepackage{graphicx}
\usepackage{dcolumn}
\usepackage{bm}
\usepackage{amsmath}

\usepackage{color}

\def\nar{\ref@jnl{New A Rev.}}          

\def\msun{M_\odot}

\begin{document}

\title[Localizations]{Facilitating follow-up of LIGO-Virgo events using rapid
  sky localization}

\author{Hsin-Yu Chen$^1$ and Daniel E. Holz$^2$}
\affiliation{$^1$Department of Astronomy and Astrophysics\\{and Kavli Institute
for Cosmological Physics}\\University of
  Chicago, Chicago, IL 60637\\
$^2$Enrico Fermi Institute, Department of Physics, and Kavli Institute
for Cosmological Physics\\University of Chicago, Chicago, IL 60637}

\begin{abstract}
Fast and effective localization of gravitational wave (GW) events could
play a crucial role in identifying possible electromagnetic
counterparts, and thereby help usher in an era of GW multi-messenger
astronomy. We discuss an algorithm for accurate and very low latency
($<$ 1 second) localization of GW sources using only the relative
times of arrival, relative phases, and relative signal-to-noise ratios
for pairs of detectors. The algorithm is
independent of distances and masses to leading order, and can be generalized to
all discrete sources detected by
ground-based detector networks.
Our approach, while developed independently, is similar to that of
BAYESTAR with a few modifications in the algorithm which result in
increased computational efficiency.
For the LIGO two detector configuration
(Hanford+Livingston) expected in late 2015 we find a median 50\% (90\%)
localization of 143 deg$^2$ (558 deg$^2$) for binary neutron stars
(for network SNR threshold
of 12, corresponding to a horizon distance of $\sim 130$ Mpc), consistent with
previous findings.
We explore the improvement in
localization resulting from high SNR events, finding that the loudest
out of the first 4 (or 10) events reduces the median sky localization area by
a factor of 1.9 (3.0) for the case of 2 GW detectors, and 2.2 (4.0) for 3 detectors.
We consider the case of multi-messenger joint detections in both the
GW and the electromagnetic (EM) spectra.
We specifically explore the case
of independent, and possibly highly uncertain, localizations, showing that the joint localization
area is significantly reduced.
We also show that a prior on the binary inclination, potentially arising from
GRB observations, has a negligible effect on GW localization.
Our algorithm is simple, fast, and accurate, and may be of particular utility in the
development of multi-messenger astronomy.

\end{abstract}

\maketitle

\section{\label{sec:intro}Introduction}

One of the most exciting potential scientific payoffs of the coming age of
gravitational wave detections is multi-messenger astronomy: the observation of a
source in both the gravitational wave (GW) and electromagnetic (EM) spectrum. Although in some cases
the EM sources may naturally be observed in coincidence with the GW signals
(e.g., {short} gamma-ray bursts [GRBs]), the more common case is expected to consist of GW 
detection followed by EM exploration of the GW sky localization error
boxes~\citep{2012ApJ...746...48M,2013PhRvL.111r1101C}. The most promising
sources for ground-based GW 
instruments are stellar mass binary inspirals and mergers of neutron stars
and/or black holes. It is not known if these systems are accompanied by EM
signatures, although there is growing evidence that short GRBs result from
binary coalescence~\citep{2010MNRAS.404..963K,2013ApJ...769...56F,2013ApJ...776...18F}. These
systems may also produce X-ray, radio, optical, and/or infrared afterglows
(e.g, kilonovae powered
by the radioactive decay of r-process elements~\citep{2013ApJ...774L..23B,2013Natur.500..547T,1999ApJ...525L.121F,2010MNRAS.406.2650M}). There will be tremendous interest in
identifying possible counterparts to LIGO sources. The success of these endeavors
will depend crucially on the ability of the LIGO network to quickly and
accurately localize GW sources on the sky.

The advanced LIGO~\citep{2010CQGra..27h4006H} detectors are currently under development, and their first
science-quality data are expected to arrive in Fall 2015~\citep{2013arXiv1304.0670L}. The
envisioned operational schedule consists of an initial GW network composed of
the two LIGO detectors (Hanford [H] and Livingston [L]), with the Virgo [V]
detector~\citep{virgotech} joining the network at a later date and at lower relative sensitivity.
GW source localization mainly relies on triangulation based on the arrival times
of the GWs at the different sites of the detector network, as has been discussed
in~\citet{2009NJPh...11l3006F,2011CQGra..28j5021F}.  The ability to triangulate a
source is crucially dependent upon the number, bandwidths, and separations of detectors in
the GW network, as well as the signal-to-noise ratio (SNR) within each detector.

There are two basic approaches to localizing a GW source. The first, exemplified
by the BAYESTAR pipeline~\citep{2015arXiv150803634S} in LALInference, utilizes the output of the LIGO detection pipeline to
directly estimate possible sky
localizations~\citep{2014ApJ...789L...5K,2014PhRvD..89h4060S,2014ApJ...795..105S}. The
detection pipeline produces rapid estimates of such quantities as time of
arrival, phase, SNR, and chirp mass, but with fairly broad uncertainties.
An alternative method to determine sky localization is to run the full parameter
estimation pipeline within LALInference~\citep{2015PhRvD..91d2003V,2013PhRvD..88f2001A}. This entails
fitting for the raw ``data $+$ noise'', utilizing Markov Chain Monte Carlo
(MCMC) and nested sampling techniques to
estimate the relevant parameters of the binary. Although this approach generates
the most reliable and tightly constrained values of parameters, including the
best estimates of the sky position, it requires significant computational resources and can take {hours},
if not {days} or longer, to run.
The performance of BAYESTAR and full parameter estimate sampling methods are comparable for GW networks
composed of two interferometers, while for three or more detectors the full
parameter estimation pipelines are superior~\citep{2014ApJ...795..105S}. This is
because, as currently configured, the detection pipeline does not produce any
information about the Virgo data if that detection is sub-threshold {($\rho\lesssim4$)}.

In this paper we develop an extremely low
latency ($<1\ \mbox{second}$ after detection) localization algorithm
incorporating the timing, {phase,} and SNR output from the detection pipeline to
estimate the localization area.  We take as input the relative timing and
{phase,} and the SNR
ratios of the GW detectors. From these parameters we estimate solely the localization area,
marginalizing over all other parameters.
Our approach is very similar to BAYESTAR~\citep{2015arXiv150803634S,2014ApJ...795..105S}, with a number of important differences:
First, our sky localization algorithm utilizes the log of the SNR
ratios, rather than the value of SNR at each detector. This obviates the need to
infer the binary chirp mass or distance, saving the time (and potential errors)
of fitting for these parameters, only to marginalize over them afterwards.
By taking the logarithm, the Gaussianity of the {likelihood} is approximately preserved.
Second, we calculate the sky location prior from the antenna power patterns and
the relative sensitivities of the detectors.  This prior applies to all discrete sources 
detected by ground-based detector networks, and can be pre-calculated in advance of
utilization of the pipeline.
Finally, our algorithm explicitly marginalizes over
only two unknown parameters: the binary orientation and inclination. Since this
is a small parameter space, we can pre-grid the entire sky and pre-calculate this
marginalization, allowing our localization algorithm to consist
primarily of lookup tables and summation, rendering it simple and
fast. Our alogrithm can be run in {$<1\,\mbox{second}$} on a laptop,
offering the potential for real-time localization as the GW source evolves
within the detectors. Since our method has been developed completely
independently of the LIGO code efforts, it also provides a
consistency check of the existing LIGO codebase (and in particular BAYESTAR).


We use our localization algorithm to explore issues related to EM follow-up. For
example, we consider the dependence of the localization area as a function of
SNR of the events, explicitly determining the expected improvement in
localization from high-SNR events. The incidence of these high-SNR events can be
predicted analytically, and depends solely on the measured event rate~\citep{2014arXiv1409.0522C}.
We show that EM priors on the inclination do not significantly help GW
localization. We also discuss the potential of cross-correlating LIGO sources with other all-sky
observatories, such as Fermi GBM, to dramatically improve sky localization.

\section{\label{sec:loc} Low Latency Localization Algorithm}

From Bayes' Theorem, the probability of a source being located at a sky position $(\theta,\phi)$,
given a set of observables $\vec{\epsilon}$, can be written as a Bayesian posterior:
\begin{equation}
\label{eqn:bayes}
f(\theta,\phi | \vec{\epsilon})=\frac{f(\vec{\epsilon}| \theta,\phi)\,f(\theta,\phi)}{f(\vec{\epsilon})}.
\end{equation}
The prior on source location can be estimated from the detector network's
antenna power pattern~\citep{2009LRR....12....2S,2011CQGra..28l5023S},
\begin{equation}
\label{eqn:pattern}
\Omega (\theta,\phi,\iota,\psi)=F_+^2(\theta,\phi,\psi)(1+\rm{cos}^2\,\iota)^2+4F^2_{\times}(\theta,\phi,\psi) \rm{cos}^2\,\iota,
\end{equation}
and sensitivity,
\begin{equation}
\label{eqn:i7}
I_7=\displaystyle\int \frac{f^{-7/3}}{S_h(f)}\,df,
\end{equation}
where $\psi$ is the orientation of the
binary within the plane of sky, $\iota$ is the inclination angle between the binary's rotation axis and the line of
sight, and $S_h(f)$ is the detector's power spectral density. The prior can be calculated from: 
\begin{equation}
\label{eqn:skyprior}
f(\theta,\phi)=\displaystyle\int f(\theta,\phi,\iota,\psi)\,d\iota d\psi,
\end{equation}
where
\begin{equation}
\label{eqn:patternprob} 
 f(\theta,\phi,\iota,\psi) \propto  \left (\sum_{i}\Omega_i(\theta,\phi,\iota,\psi)\,I_{7,i}\right )^{3/2} \,\rm{sin}\,\theta \, \rm{sin}\,\iota,
\end{equation}
and where the sum goes over each detector in the network ($i=$H, L,
V,$\,\ldots$). The likelihood function is expressed as a function of the sky
location, $(\theta,\phi)$,  and the GW observables, $\vec{\epsilon}$:
\begin{equation}
\label{eqn:likelihood}
f(\vec{\epsilon}|\theta,\phi)\propto \rm{exp}\left(-\Delta \chi^2_{\vec{\epsilon}}(\theta,\phi)/2\right)
\end{equation}
where $\Delta \chi^2_{\vec{\epsilon}}$ can be calculated from a chi-square minimization process:
$\Delta \chi^2_{\vec{\epsilon}}(\theta,\phi)=\chi^2_{\vec{\epsilon}}(\theta,\phi)-\chi^2_{\vec{\epsilon},{\rm min}}(\theta_{\rm min},\phi_{\rm min})$.
For our algorithm the observables are the arrival times of the signal, $t_i$, the phase, $\eta_i$, and
the SNR, $\rho_i$, measured at each detector.
If the SNR measurements are independent of the arrival times and phases, the chi-square
expression can be separated:
$\chi^2_{\vec{\epsilon}}(\theta,\phi)=\chi^2_{\zeta}(\theta,\phi)+\chi^2_{\rho}(\theta,\phi),$
where $\zeta$ represents the contribution from arrival time and phase, and
$\rho$ represents the contribution from SNR (see below).
We assume the errors in arrival time, phase, and SNR are Gaussian
and independent between detectors. This appears to be a fair assumption~\citep{2015ApJ...804..114B},
 although our method can be generalized if the error properties
are known.

The differences in the arrival time between detectors allows us to triangulate for
the sky position of the source. Using only this information, every pair of detectors will 
localize the source to an annulus on the sky. Three detectors with two
intersecting annuli will localize the source to two intersection regions, while four or more
detectors can localize the source to a single distinct region. This technique has been
discussed by~\citet{2009NJPh...11l3006F,2011CQGra..28j5021F} using a flat-sky 
approximation; we rederive these results in full generality.

The phase of a binary, $\eta$, at time $t$, measured in the $i$th
detector is a function of 
the detector orientation, and the binary sky location, inclination, and orientation: $\eta (t)+\eta_i(\theta,\phi,\iota,\psi)$. 
Following ~\citet{2009LRR....12....2S}, we find:
$$\eta_i(\theta,\phi,\iota,\psi)={\rm tan}^{-1}\,\left (\frac{2F_{\times,i}(\theta,\phi,\psi) \rm{cos}\,\iota}{F_{+,i}(\theta,\phi,\psi)(1+\rm{cos}^2\,\iota)}\right ).$$
By considering only the phase difference between detectors, we avoid needing to
determine the absolute phase of the binary. {The
sky location remains degenerate with the inclination and orientation, and this
can be subsequently marginalized over.
There are additional phase differences due to the separation of the detectors on
Earth; this can be inferred from the difference in signal arrival times. 
The phase and timing measurements are therefore correlated.}

We follow the approach of~\citet{2014PhRvD..89d2004G,2009NJPh...11l3006F,2011CQGra..28j5021F} to
estimate the timing and phase covariance matrix.
These errors depend on the SNR and the detector effective bandwidth of the source, $\sigma_f^2=\bar{f^2}-({\bar{f}})^2$:
\begin{equation}
\label{eqn:terr}
{\rm cov}(t,\eta)\equiv
\begin{bmatrix}
  \sigma^2_{tt} & \sigma^2_{t \eta}    \\
  \sigma^2_{\eta t}  & \sigma^2_{\eta \eta}  
 \end{bmatrix}=
\begin{bmatrix}
  \frac{1}{(2\pi \rho\sigma_f)^2} & -\frac{\bar{f}}{2\pi\rho^2\sigma_f^2}   \\
  -\frac{\bar{f}}{2\pi\rho^2\sigma_f^2} & \frac{\bar{f^2}}{(\rho\sigma_f)^2}  
 \end{bmatrix},
\end{equation}
where $\bar{f^n}=4 \displaystyle\int\frac{|\bar{h}(f)|^2}{S_h(f)}f^n \,df.$ 
The error in \emph{measured}\/ arrival times and phases are expected to be
distributed as a 2D Gaussian centered at the true value with covariance
given by Eq.~\ref{eqn:terr}. In reality,
the simulated binary neutron stars merger signals in ~\citet{2014ApJ...795..105S} 
follow a Gaussian distribution with a standard deviation that is 1.4 and 1.3 times larger in arrival time and 
phase, respectively. The
difference in arrival times and phase between pairs of detectors is a linear function of
the independent observables, $t_i$ and $\eta_i$:
\begin{align}
\nonumber\Delta t_{ij}&\equiv t_i-t_j, \\
\nonumber\Delta \eta_{ij}&\equiv \eta_i-\eta_j, \;\;\; i,j={\rm H,L,V,} \ldots\, \ \mbox{with}\  i\neq j.
\end{align}
Thus the \emph{measured}\/ timing and phase differences can still be described by 2D Gaussians.
If there are $N$ detectors in a given GW network, there are $N-1$ independent
timing and phase differences. Without loss of generality  
we can denote one of the detectors as reference detector \emph{I}, and define
all time-of-arrival and phase differences with respect to this fiducial detector: 
\begin{align}
\label{eqn:delt}
\nonumber \Delta t_i&\equiv t_{I}-t_i, \\
\Delta \eta_i&\equiv \eta_{I}-\eta_i,\;\;\; i\neq I.
\end{align}
The chi-square value is:
\begin{equation}
\label{eqn:chi2t}
\chi^2_{\zeta}(\theta,\phi,\iota,\psi)=\displaystyle\sum_{i,j} \zeta_i^{T}(\theta,\phi,\iota,\psi) V^{-1}_{ij} \zeta_j(\theta,\phi,\iota,\psi)
\end{equation}
where 
$$\zeta_i(\theta,\phi,\iota,\psi)\equiv \begin{bmatrix} \Delta {t_i}-\Delta {t_i}(\theta,\phi) \\ \Delta {\eta_i}-\Delta {\eta_i}(\theta,\phi,\iota,\psi)\end{bmatrix}$$
and $V_{ij}$ is the covariance matrix.
Since all the differences are in reference to the same detector $I$, any two differences are not independent.

Following Eq.~\ref{eqn:terr}, we find that when $i=j$ the components of $V_{ij}$
are given by~\footnote{If variables are correlated, the variance of their 
sum is the sum of their covariances.}:

\begin{align}
\nonumber \sigma^2_{\Delta t_i\Delta t_i}&=\sigma_{t_{\rm I}t_{\rm I}}^2+\sigma_{t_it_i}^2,\\
\nonumber \sigma^2_{\Delta \eta_i\Delta \eta_i}&=\sigma_{\eta_{\rm I}\eta_{\rm I}}^2+\sigma_{\eta_i\eta_i}^2,\\
\nonumber \sigma^2_{\Delta t_i\Delta \eta_i}&=\sigma^2_{\Delta \eta_i\Delta t_i}=\sigma_{t_{\rm I}\eta_{\rm I}}^2+\sigma_{t_i\eta_i}^2.
\end{align}

If $i\neq j$,

\begin{align}
\nonumber \sigma^2_{\Delta t_i\Delta t_j}&=\sigma_{t_{\rm I}t_{\rm I}}^2,\\
\nonumber \sigma^2_{\Delta \eta_i\Delta \eta_j}&=\sigma_{\eta_{\rm I}\eta_{\rm I}}^2,\\
 \sigma^2_{\Delta t_i\Delta \eta_j}&= \sigma^2_{\Delta \eta_i\Delta t_j}=\sigma_{t_{\rm I}\eta_{\rm I}}^2.
\end{align}
We marginalize over $(\iota,\psi)$ using Eqs.~\ref{eqn:skyprior} and~\ref{eqn:patternprob}:
\begin{equation}
\label{eqn:chi2teff}
\chi^2_{\zeta}(\theta,\phi)=-2\, {\rm log} \left
(\frac{1}{f(\theta,\phi)}\displaystyle \int
e^{-\chi^2_{\zeta}(\theta,\phi,\iota,\psi)/2}\, f(\theta,\phi,\iota,\psi) \,
d\iota d\psi \right )
\end{equation}

In addition to the timing and phase differences, the relative values of SNR received by
each detector in the GW network also provides constraints on the source location.
Different detectors are sensitive to different polarizations of the incoming
waves, and the relative response in each detector depends upon 
the binary sky location $(\theta,\phi)$ and orientation $(\iota,\psi)$, 
as well as the detector placement on the Earth, orientation of the detector
arms, and the detector sensitivity, $S_h(f)$.
The SNR measured in a given GW detector depends on the antenna response of the
detector, as well as the binary source parameters, such 
as the luminosity distance, redshift, and chirp mass.
Aside from chirp mass, these binary parameters are poorly constrained by the detection
pipeline.
However, since for our purposes we are only
concerned with sky position, we can eliminate the other binary parameters by taking the
ratio of the SNR measured at pairs of detectors. This ratio is
solely a function of the detectors' antenna power patterns (Eq.~\ref{eqn:pattern}) and
sensitivities (Eq.~\ref{eqn:i7}):
$$\frac{\rho^2_ i}{\rho^2_j}(\theta,\phi,\iota,\psi)=\frac{\Omega_i(\theta,\phi,\iota,\psi)\,I_{7,i}}{\Omega_j(\theta,\phi,\iota,\psi)\,I_{7,j}}.$$
We estimate the Gaussian uncertainty in SNR following the Fisher matrix
calculation~\citep{1994PhRvD..49.2658C}:
\begin{equation}\label{eqn:amperr}
\frac{\sigma_\rho}{\rho}=\frac{1}{\rho},
\end{equation} 
which corresponds to a $\sim 10\%$
error in SNR measurement for a detection threshold of $\rho_{\rm
th}\sim12$. The {\em measured}\/ SNR will be Gaussian distributed about
the theoretical value $\tilde{\rho}=G(\rho,\sigma_{\rho})$. Unlike the arrival
time difference, the SNR ratio is not linear in SNR, so the error in the ratio
is not described by a Gaussian. However, by taking the logarithm of the SNR
ratio:
\begin{equation}
\label{eqn:ampratio}
R_{ij}(\theta,\phi,\iota,\psi)\equiv {\rm log}\,\frac{\rho_ i}{\rho_j}={\rm log}\,\rho_i-{\rm log}\,\rho_j,
\end{equation}
we find that the error in $R$ approximates a Gaussian with variance~\footnote{The error
becomes increasingly non-Gaussian when the
SNR in one detector falls below $\sim 5$. This leads to a loss of the catch
percentage at high confidence level, though this deviation is insignificant
($<1\%$).}
$$\sigma_{R_{ij}}^2=\frac{1}{\rho_i^2}+\frac{1}{\rho_j^2}.$$

\begin{figure}[tb!]
\begin{center}
\includegraphics[width=1.0\columnwidth]{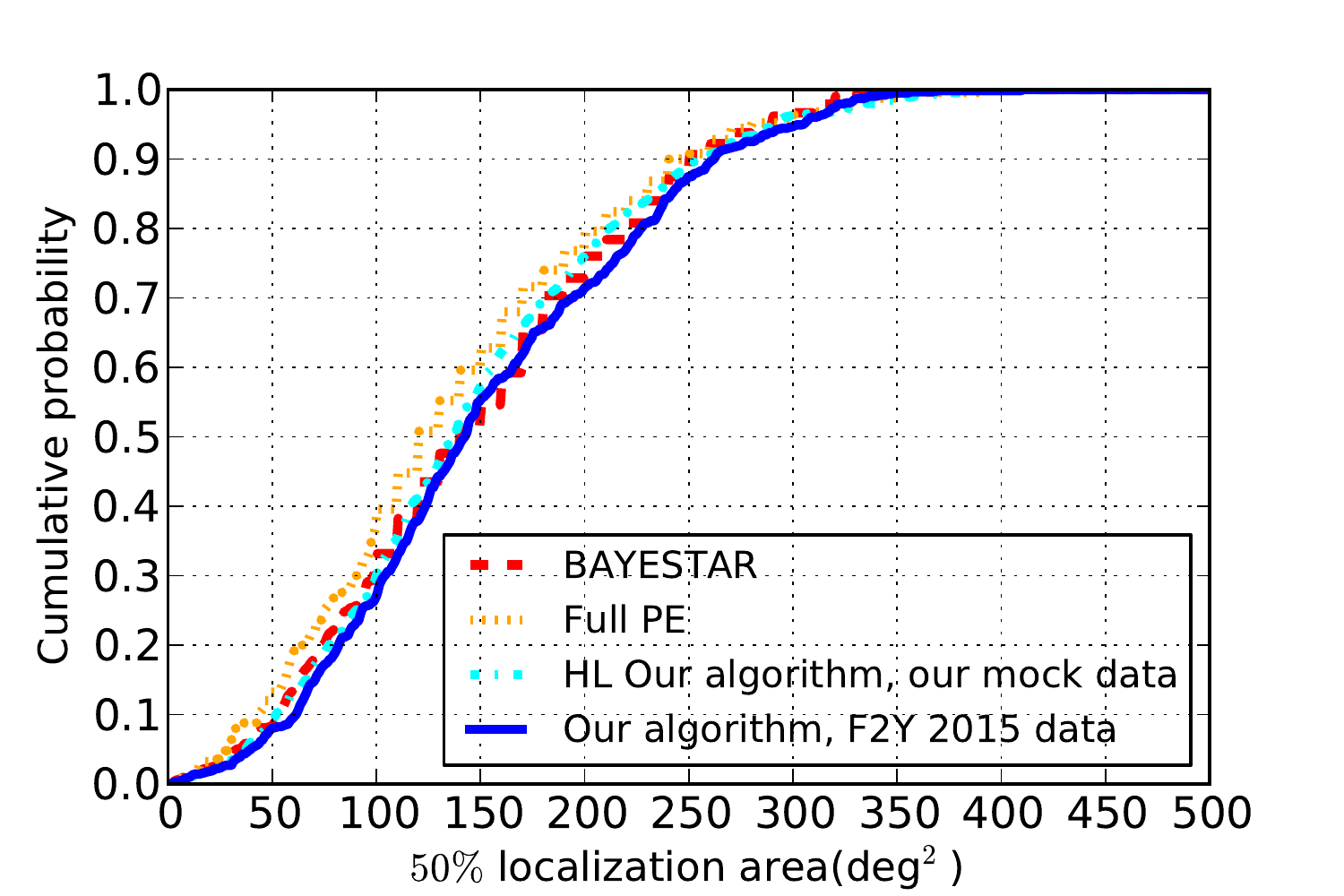}
\caption{\label{fig:hl_area}
Cumulative $50\%$ confidence localization areas of the 630 simulated sources taken from ~\citet{2014ApJ...795..105S}
 detected by LIGO Hanford+Livingston with the noise curve in
 LIGO-T1200307. 
}
\end{center}
\end{figure}

\begin{figure}[tb!]
\begin{center}
\includegraphics[width=1.0\columnwidth]{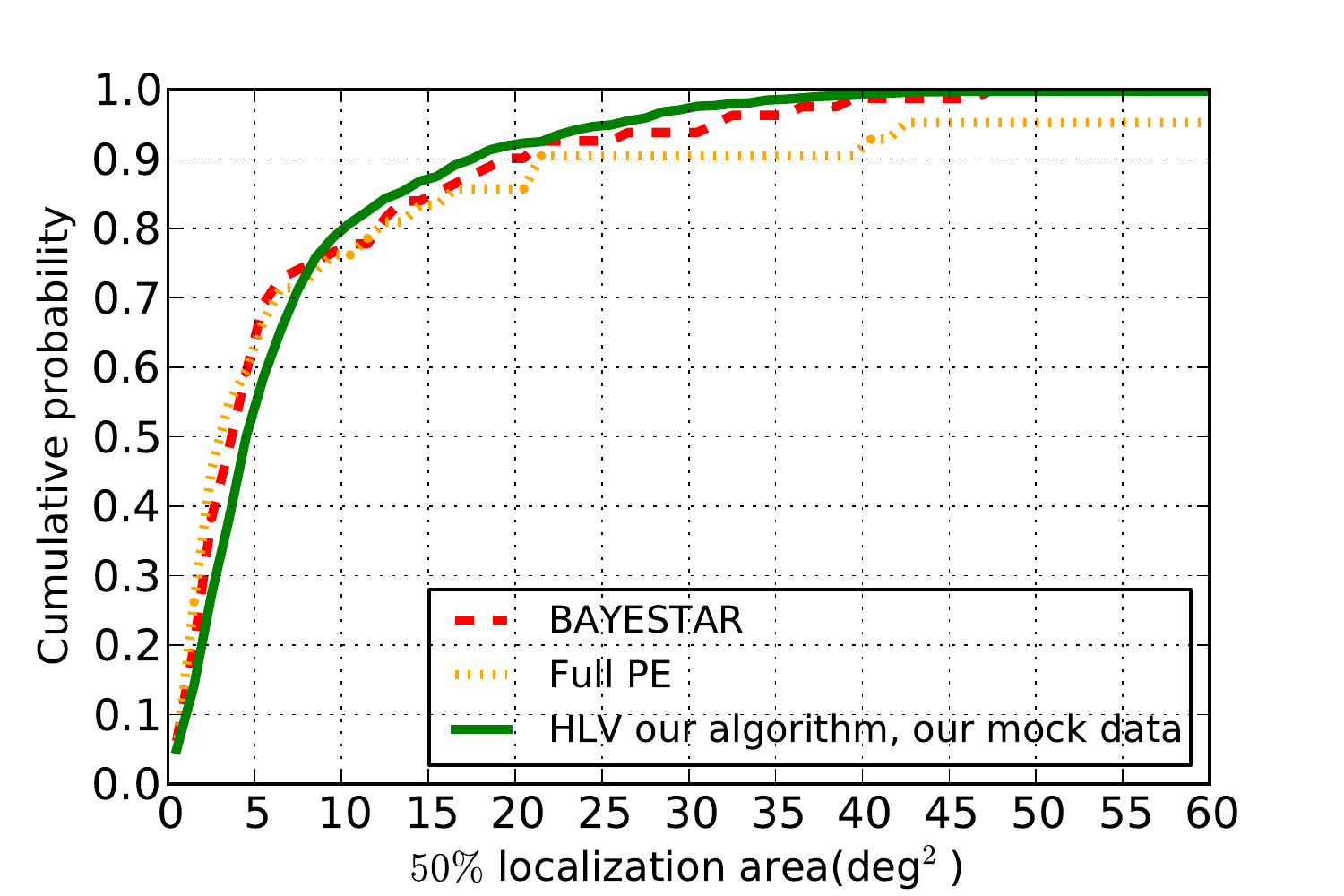}
\caption{\label{fig:hlv_area}
Cumulative $50\%$ confidence localization areas of the 81 simulated sources taken from ~\citet{2014ApJ...795..105S}
 detected by LIGO Hanford+Livingston and Virgo with the noise curves in
 LIGO-T1200307 and~\citet{2013arXiv1304.0670L}. We only
 consider events with SNR greater than 4 in all three detectors.}
\end{center}
\end{figure}

Similar to the case with arrival times, there are $N-1$ independent SNR ratios from a network consisting of $N$
independent detectors. Therefore we 
define $R_i\equiv {\rm log} (\rho_I/\rho_i),\, i\neq I$ with an arbitrary
reference detector \emph{I}. The chi-square for SNR is then:
\begin{equation}
\label{eqn:chi2r}
\chi^2_{\rho}(\theta,\phi,\iota,\psi)=\displaystyle\sum_{i,j} [\tilde{R_i}-R_i(\theta,\phi,\iota,\psi)] V^{-1}_{ij}[\tilde{R_j}-R_j(\theta,\phi,\iota,\psi)]
\end{equation}
where $V_{ij}\equiv {\rm cov}(R_i,R_j)$ is the covariance matrix. If $i=j$ then
$$\sigma_{R_iR_i}^2=\frac{1}{\rho_I^2}+\frac{1}{\rho_i^2},$$
while if $i\neq j$ then we find:
\begin{equation}
\begin{split}
\sigma^2_{R_iR_j}&=
\frac{1}{2}[\sigma_{R_i}^2+\sigma_{R_j}^2-{\rm var}(R_i-R_j)]\\
&=\frac{1}{2}(\sigma_{R_i}^2+\sigma_{R_j}^2-\sigma_{R_{ji}}^2) \\
&=\frac{1}{2}\left (\frac{1}{\rho_I^2}+\frac{1}{\rho_i^2}+\frac{1}{\rho_I^2}+\frac{1}{\rho_j^2}-\frac{1}{\rho_i^2}-\frac{1}{\rho_j^2} \right ) \\
&=\frac{1}{\rho_I^2}.
\end{split}
\end{equation}
By taking the logarithm of the power ratio (Eq.~\ref{eqn:ampratio}) we are able
to estimate the covariance matrix analytically.

As shown in Eq.~\ref{eqn:chi2r}, the directional
information is degenerate with binary
inclination and orientation. We marginalize over $(\iota,\psi)$ using
Eqs.~\ref{eqn:skyprior} and~\ref{eqn:patternprob}:

\begin{equation}
\label{eqn:chi2reff}
\chi^2_{\rho}(\theta,\phi)=-2\, {\rm log} \left
(\frac{1}{f(\theta,\phi)}\displaystyle \int
e^{-\chi^2_{\rho}(\theta,\phi,\iota,\psi)/2}\, f(\theta,\phi,\iota,\psi) \,
d\iota d\psi \right ).
\end{equation}

\begin{figure}[t]
\begin{center}
\includegraphics[width=1.0\columnwidth]{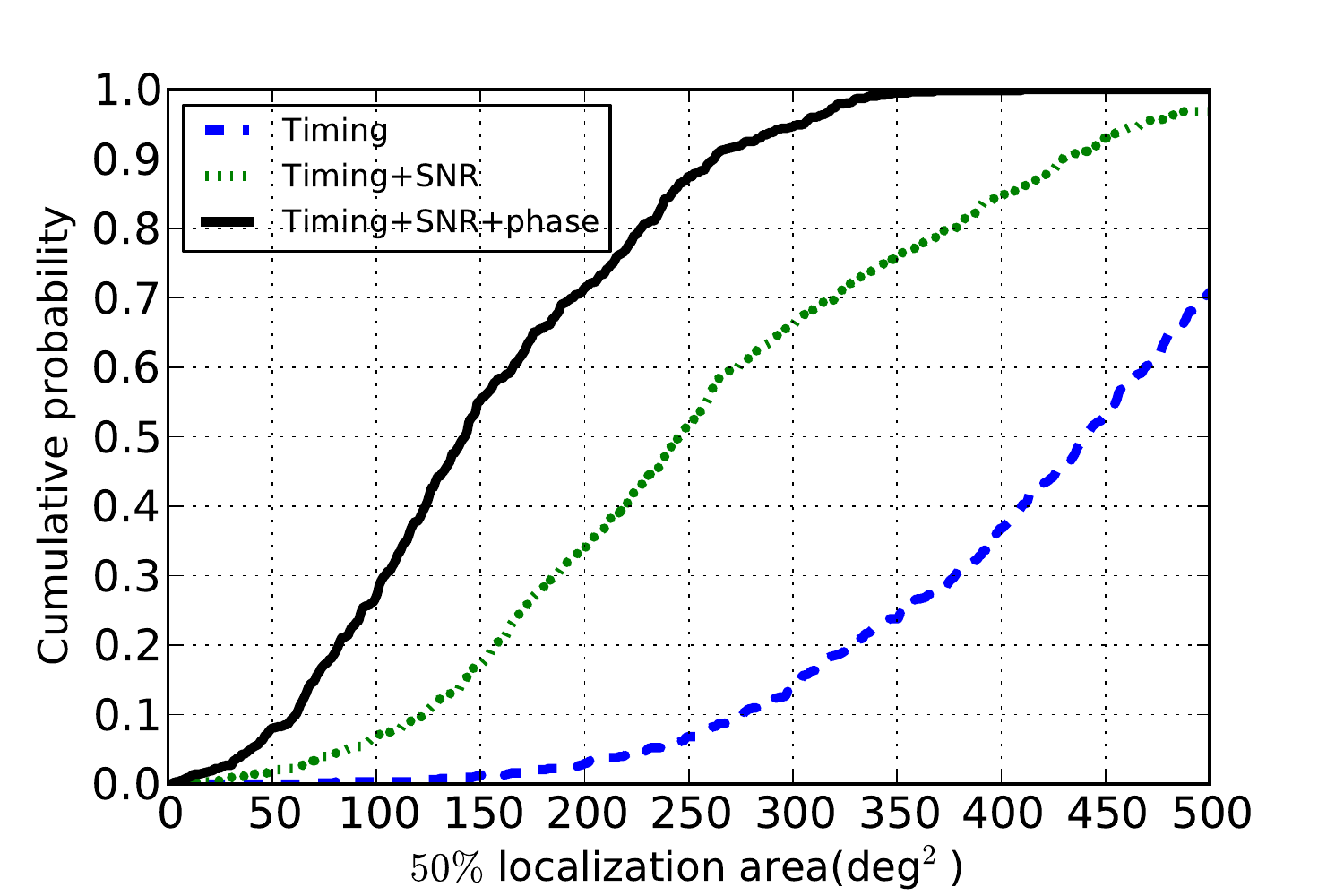}
\caption{\label{fig:addphase}
We compare the improvement of using time of arrival+SNR and time of arrival+SNR+phase.
The simulated sources were taken from ~\citet{2014ApJ...795..105S} and localized
with LIGO Hanford and Livingston.
}
\end{center}
\end{figure}

The sky prior (Eq.~\ref{eqn:skyprior}) also contributes to the posterior, so we
define the total chi-square as: 
$\chi^2(\theta,\phi) \equiv \chi^2_{\vec{\epsilon}}(\theta,\phi)-2\,{\rm
  log}\,f(\theta,\phi)$. 
We grid the sky uniformly in $({\rm cos}\,\theta,\phi)$ (or in $(\theta,\phi)$ when close to the
poles)~\footnote{We grid the sky uniformly in $({\rm cos}\,\theta,\phi)$ with $d\,{\rm cos}\,\theta=d\phi=0.005$ ($\sim$0.3 degree) 
for the two detector case. We use a finer gridding in the 3 detectors case, $d\,{\rm cos}\,\theta=d\phi=0.0025$ ($\sim$0.14 degree)}, 
calculate $\chi^2$ at each grid point, and then find the minimum
$\chi^2_{\rm min}$ from which to determine
$\Delta \chi^2$ at each point and establish the
posterior (Eq.~\ref{eqn:bayes}). We use
a {contour defined by $\Delta \chi^2=b_p$} to determine the boundary of the
localization area, such that 
\begin{equation}
\label{eqn:int_prob} 
\frac{1}{n}\displaystyle\int_{\Delta \chi^2(\theta,\phi)<b_p}e^{-\Delta \chi^2(\theta,\phi)/2}\, d\theta d\phi=p\,\%,
\end{equation}
where $n$ normalizes the probability to 1. We
vary the $\Delta \chi^2$ value of the boundary, $b_p$, until the integral in
Eq.~\ref{eqn:int_prob} reaches the desired confidence level $p\,\%$. 
If the posterior is Gaussian, we can calculate $b_p$ analytically. The sky prior
(Eq.~\ref{eqn:skyprior}) and the
marginalization (Eq.~\ref{eqn:chi2reff}) break the 
Gaussianity. However, we still use
contours of constant $\Delta \chi^2$ as the
boundaries of our confidence level, where we explicitly compute the
probability incorporated within each contour rather than assuming the analytic
value appropriate for true Gaussian distributions~\citep{Press:2007:NRE:1403886}. 

\begin{figure}[t]
\begin{center}
\includegraphics[width=1.0\columnwidth]{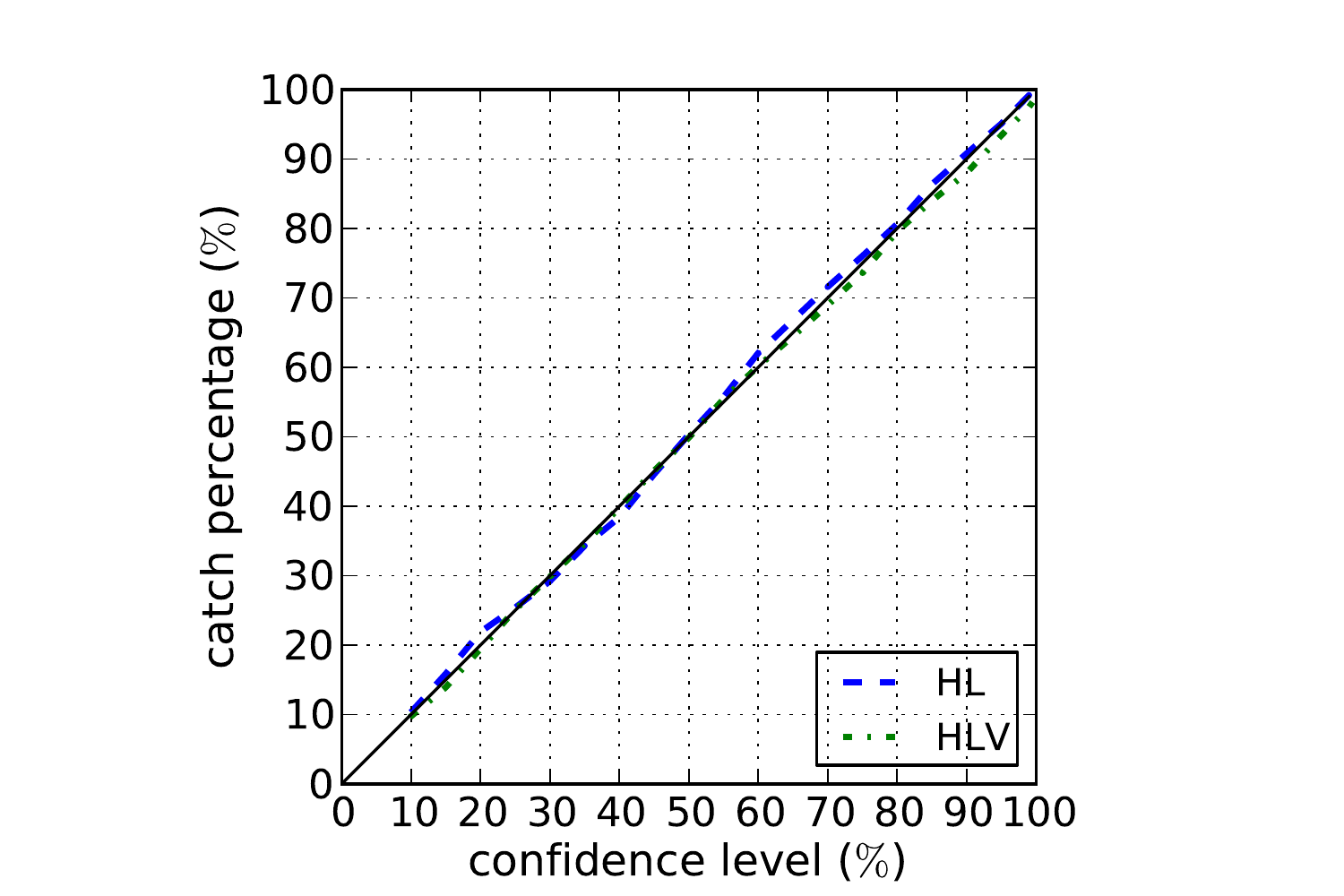}
\caption{\label{fig:fidelity} Integrated confidence level versus the actual
  captured percentage. The samples were taken from ~\citet{2014ApJ...795..105S} and localized with 
  {LIGO Hanford+Livingston (HL) and HL+Virgo (HLV)}. The black line
  is the $x=y$ line (perfect consistency). We find that the fraction of binaries
  found within our sky localization contours is in very good agreement with the predicted
  percentage.}
\end{center}
\end{figure}

In summary, for a given event we use Eqs.~\ref{eqn:bayes},~\ref{eqn:likelihood},
and~\ref{eqn:int_prob} to estimate the localization area on the sky. As a test,
we apply our localization algorithm to the simulated source catalog from~\citet{2014ApJ...795..105S}, 
and compare our localization areas to those found from the BAYESTAR
algorithm. As shown in Fig.~\ref{fig:hl_area}, we find
excellent consistency.
We also explore the contribution to the localization determination arising
from incorporating SNR and phase information in addition to timing, finding that the resulting 
localization areas decrease significantly (Fig.~\ref{fig:addphase}).
In order to test our Bayesian posterior, we {determine the fraction of
binaries which fall within our localization areas}, 
and compare this to the predicted percentages.
Fig.~\ref{fig:fidelity} shows that our formalism
is {self consistent}.
{In order to further study the properties of our low-latency localization
  algorithm, we simulate a population of binary neutron star. We follow the
  Monte Carlo approach we presented in~\citet{2014arXiv1409.0522C}, utilizing the
  same noise curve as~\citet{2014ApJ...795..105S}.}
For each binary we calculate the true SNR and time-of-arrival in each
detector. We then
add in Gaussian noise for each of these quantities, mimicking the actual
measurements (ignoring glitches; we have assumed a threshold SNR of
$\rho_{\rm th}=12$ to mimimize the importance of these). 
{Our resulting  distribution of localization areas is comparable to that in
Figs.~\ref{fig:hl_area} and~\ref{fig:hlv_area}, indicating that our Monte Carlo sources and pipelines
are consistent with that of~\citet{2014ApJ...795..105S}.}

\section{Loudest Event}\label{sec:loudest}
\begin{figure}
\centering 
\includegraphics[width=1.0\columnwidth]{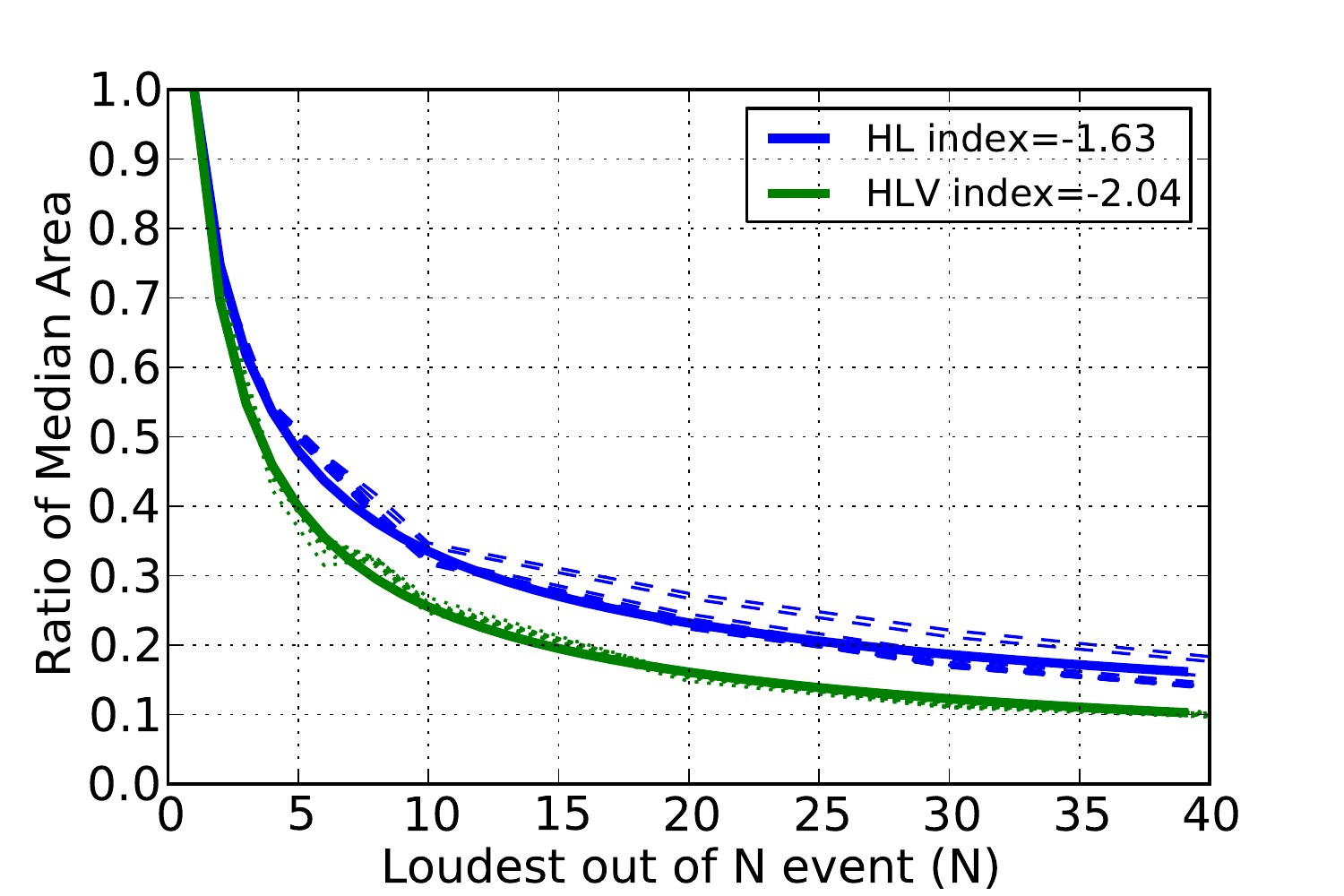}
\caption{\label{fig:arearatio}
Ratio between the median localization area for the loudest event out of $N$ detections and that for all
  events ($N=1$). The SNR of the loudest events can be estimated from Eq.~\ref{eq:loudest}. 
  For 3 detectors (HLV) the areas roughly scale as SNR$^{-2}$ (index=-2.04), while for 2 detectors (HL) 
  the scaling is shallower (index=-1.63).  
}
\end{figure}

As shown in Fig.~\ref{fig:hl_area} and ~\ref{fig:hlv_area}, there is a broad distribution of localization
areas for any given set of sources. This is related to the broad distribution of
measured SNRs of the sources, which is in turn related to the distribution of
distances to the sources.
If not all GW triggers can be followed, the louder (higher SNR) events offer the
opportunity for better localization and more efficient follow-up.
In Fig.~\ref{fig:arearatio} we show the median localization area of the loudest
source out of $N$ events. The $N=1$ case corresponds to the full distribution of
localization areas for all events (also seen in Fig.~\ref{fig:hl_area}).
From Fisher matrix calculations the localization area is expected to scale as Area$\sim {\rm SNR}^{-2}$. 
This is true for 3 or more detectors, while for 2 detectors the sky locations
are generally poorly constrained within a broad timing ring and the localization
area will scale more slowly with improving SNR. In our loudest event
paper~\citep{2014arXiv1409.0522C} we show that the distribution of detected
SNRs follows a universal distribution. We also
derive the universal loudest event distribution, which depends upon  
the detection threshold, $\rho_{\rm th}$, and the number of events, $N$. The
cumulative probability is: 
\begin{equation}\label{eq:loudest}
F_{\rho_{\rm max}}=\left ( 1-\left (\frac{\rho_{\rm th}}{\rho_{\rm max}} \right )^3 \right )^N.
\end{equation}
In \citep{2014arXiv1409.0522C} we pointed out that Eq.~\ref{eq:loudest} 
is independent of the detector network, detector sensitivity, and all properties
of the sources.
Eq.~\ref{eq:loudest} can applied to all situations (e.g. 2 or 3 detectors
operating at different sensitivities looking for BNS, NS-BH, or supernovae),
and we can predict the highest SNR out of $N$ detections
  analytically. 
For example, the median of this distribution is given by:
$$\rho_{\rm max}=\frac{\rho_{\rm th}}{(1-0.5^{1/N})^{1/3}}.$$
If the localization area scales as Area$\sim {\rm SNR}^{-2}$, we would expect the ratio between the 
median area of the loudest out of $N$ events, $A_{\rm max\,N}$, and the median
area of all events ($N=1$), $A_1$, to scale as:
\begin{equation}\label{eq:ratio}
\frac{A_{\rm max\,N}}{A_1}=\left (\frac{\rho_{\rm max\, N}}{\rho_1}\right )^{-2}=\left(\frac{1-0.5^{1/N}}{0.5}\right )^{2/3}.
\end{equation}
We fit for the area-SNR scaling index in Fig.~\ref{fig:arearatio}. 
For 3 detectors we see that the area scales as SNR to the power of -2.04, while for
the 2 detector case the scaling is shallower (-1.63) as expected.  

Eq.~\ref{eq:ratio} shows that the localization area shrinks rapidly as
additional binaries are initially added: waiting for the loudest out of the first 4 events will
reduce the expected sky area required for follow-up by a factor of
1.9 (2 detectors) and 2.2 (3 detectors). This improvement declines for larger sets of binaries; going from  
4 to 10 binaries reduces the sky area by an additional factor of 1.6 (2 detectors) and 1.8 (3 detectors).
This suggests that there may be an optimal strategy for selecting events for EM
follow-up, based on the expected event rate and the observational constraints of
the follow-up facility.

\section{GRB Joint Localizations}\label{sec:fermi}
Short GRBs are thought to be the very bright EM counterpart of binary
  neutron star or neutron star-black hole coalescence~\citep{2010MNRAS.404..963K,2013ApJ...769...56F,2013ApJ...776...18F}. 
  If a short GRB is accompanied by GW emission, joint GW+EM detection would
confirm the binary nature of the short
GRB engine as well as help constrain the binary parameters. For example, the
existence of gamma-ray emission would suggest that at least one of the members of the binary
is a neutron star, and measurements of the inclination would help constrain
properties of the beamed jet. In addition, the shape and area of the
localization of the GRB in gamma-rays is
independent of the Advanced LIGO \& Virgo localization, and thus combining the
two localization areas can significantly constrain the sky position of the source.

{We would like to explore the benefit to sky localization of a prior
  constraint on the inclination.}
{Let us consider the case where a GRB has been observed, and we thereby
  have a prior on the GRB beaming angle which comes into
  Eqs.~\ref{eqn:skyprior},~\ref{eqn:chi2teff}, and~\ref{eqn:chi2reff}.}
We randomly generate 3,000 binary neutron star mergers
following the approach in~\citet{2014arXiv1409.0522C}. As suggested by observations, we assume
that GRBs are beamed
within the range
$1$--$10^{\circ}$~\citep{2006ApJ...650..261S,2006ApJ...653..468B,Coward:2012uz,Fong:2012wz},
and we confine the orientation of our binaries to be uniformly distributed in ${\rm
cos}\,\iota$ {within} $\pm{\rm cos}\,1^{\circ},\; \pm{\rm
    cos}\,5^{\circ}, \; \pm{\rm cos}\,10^{\circ}$ {(3 samples of 1,000 binaries each)}. 
For each beaming angle distribution, we localize the source by 
marginalizing over the corresponding range of inclination angles 
(``$\sin\iota\leq \sin\theta_j$'' which allows for both face-on and tail-on
beaming; see Table~\ref{table:beaming}) or 
over the full range of inclinations (``random $\iota$''). We then compare the
localization area and summarize the results 
in Table~\ref{table:beaming}. {Using the beaming prior only improves the
localization by $<10\%$, even in the case of $1^\circ$ beaming.}
 This is due to the unknown direction of rotation; a
  clockwise binary produces a different phase pattern than a counter-clockwise
  binary, and knowing that a binary is almost face-on {\em or}\/ almost tail-on
  is not as useful as knowing that a binary is definitely one or the other. If
  we assume that we break the clockwise/counter clockwise degeneracy, the
  localization improves significantly, approaching the ``known
  $\iota$'' case. It is to be noted that additional measurements of phase, e.g.,
from Virgo, LIGO-India, or KAGRA, do not help to break this degeneracy.
This is because a clockwise binary ``above'' a detector produces an identical
waveform to a counter-clockwise binary ``below''. Alternatively, if the sky
position is known, the GW network can determine whether the binary is orbiting
clockwise or counter-clockwise.
\begin{table}[tp]%
\centering
\begin{tabular}{clccc}
\toprule
\hline\hline
       $\theta_j$     & known $\iota$  	& $\sin\iota\leq \sin\theta_j$	&random $\iota$		\\ \midrule
\hline
        1$^{\circ}$  			& 340 deg$^2$    	& 490 deg$^2$	&540 deg$^2$							\\ 
        5$^{\circ}$  			& 340 deg$^2$    	& 500 deg$^2$	&530 deg$^2$						\\ 
        10$^{\circ}$			& 350 deg$^2$    	& 510 deg$^2$	&530 deg$^2$							\\
        10$^{\circ}$, clockwise			& \hspace*{0.4cm}$\cdots$
& 350 deg$^2$	& \hspace*{0.1cm}$\cdots$ 							\\  
\hline\hline
\end{tabular}
\caption{Hanford+Livingston median 90$\%$ localization area for binary
  neutron stars beamed within $\theta_j$.  The localizations are for three
  different sets of priors: knowing the exact inclination (``known $\iota$''), knowing the beaming
  angle (``$\sin\iota\leq \sin\theta_j$'') but not whether it is clockwise or
  counter-clockwise, and assuming no information about beaming (``random
  $\iota$'').
The last row corresponds to binaries that are beamed
within $10^\circ$ and revolve in the clockwise direction on the sky, and where
the prior assumes not only the beaming but also the clockwise direction.
\label{table:beaming}}
\end{table}

{Although a beaming prior is of limited utility in determining position, 
the contribution of an independent localization from EM measurements of a GRB
can be significant.}
The Fermi Gamma-ray Burst Monitor (GBM) telescope has a field-of-view of
8~steradians. It observes $\sim 45$ short GRBs a year, potentially serving as a
triggering system for binary mergers. A limitation of
the Fermi GBM is that it can only localize sources within tens to hundreds of
square degrees~\citep{2009AIPC.1133...40B}. We conservatively assume that Fermi GBM
localizes sources to a 100 deg$^2$ ($1\sigma$) 2D circular Gaussian. This independent
likelihood function contributes to the posterior as an additional $\chi^2_{\rm
Fermi}$.
Fig.~\ref{fig:fermi} shows the significant improvement in source localization
which results from joint detections from Fermi and advanced GW detectors. The
median of the 90\% confidence ($\sim2\sigma$)  
area for joint detection lies around 120~deg$^2$ even with only two GW detectors, {as
  compared with 460~deg$^2$ and 560~deg$^2$ for the GRB and GW localizations, respectively.}
In this case the beaming priors make no difference whatsoever in the localization.

\section{Discussion}\label{sec:discussion}
We have developed a low latency localization algorithm for ground-based GW
detector networks. Our approach requires the timing, phase, and SNR of the
sources (provided by the detection pipeline), and within a fraction of a second
produces a 2D localization.
We have implemented our algorithm for the case of non-spinning compact binary
coalescences, assuming representative output from the detection pipeline
(GSTLAL); this can in principle be generalized to other (preferably modeled) burst sources.
{Although the sky location is degenerate with other physical properties
(e.g. masses, distances, and spins of the binaries), our algorithm avoids the
need to estimate these paramters by instead utilizing the difference in arrival
times, difference in phases, and the ratio of measured SNRs.}
This increases the simplicity and speed of localization, while reducing 
potential errors.
Imprecise waveforms can result in errors in the SNR measurements, but the ratio
of SNRs between detectors is less sensitive to these errors.  By taking the SNR
ratio, the parameter space is highly reduced.  For binary mergers, the two
relevant unknowns are the binary orientation and inclination; it is
computationally inexpensive to pre-calculate the marginalization of these over
the entire sky~\footnote{We grid the binary inclination and orientation into 50
  points each (2,500 points in total), and calculate the marginalization at each
  grid point.  We assume that the noise curve is known and fixed, allowing
  pre-calculation of the $I_7$ term. Changes in the noise curves can be captured
  by the detection pipeline, and incorporated into our algorithm.  The ratio of
  $I_7$ terms, which is the relevant quantity since we are concerned with the
  ratio of SNR, is insensitive to changing total mass for
  $M\lesssim10\,\msun$.}. The timing and sky priors can also be pre-gridded and
pre-calculated, simplifying the localization algorithm to a $\chi^2$
{determination using a look-up table combined with direct summation}. Once the
tables are loaded into memory, the entire algorithm takes $<1$ second on a
laptop (using a Python 2 script running on one thread on a 2.3 Ghz Intel Core
i7). It would be straightforward to parallelize the algorithm (e.g., by having
each core analyze a fraction of the sky), leading to sky localization within 0.1 seconds
or better.
 There is ongoing work to speed up the detection pipeline
~\citep{fastdetection}, including
the possibility of event detection before merger. This capability is only of use
when coupled with rapid localization; our prompt and accurate algorithm offers
the possibility of real-time localization contemporaneous with the inspiral and
merger of the binary.
It is to be noted that our algorithm can be generalized to calculate distance as
well as sky position, allowing for the possibility of low-latency 3D localization.

\begin{figure}[t]
\begin{center}
\includegraphics[width=1.0\columnwidth]{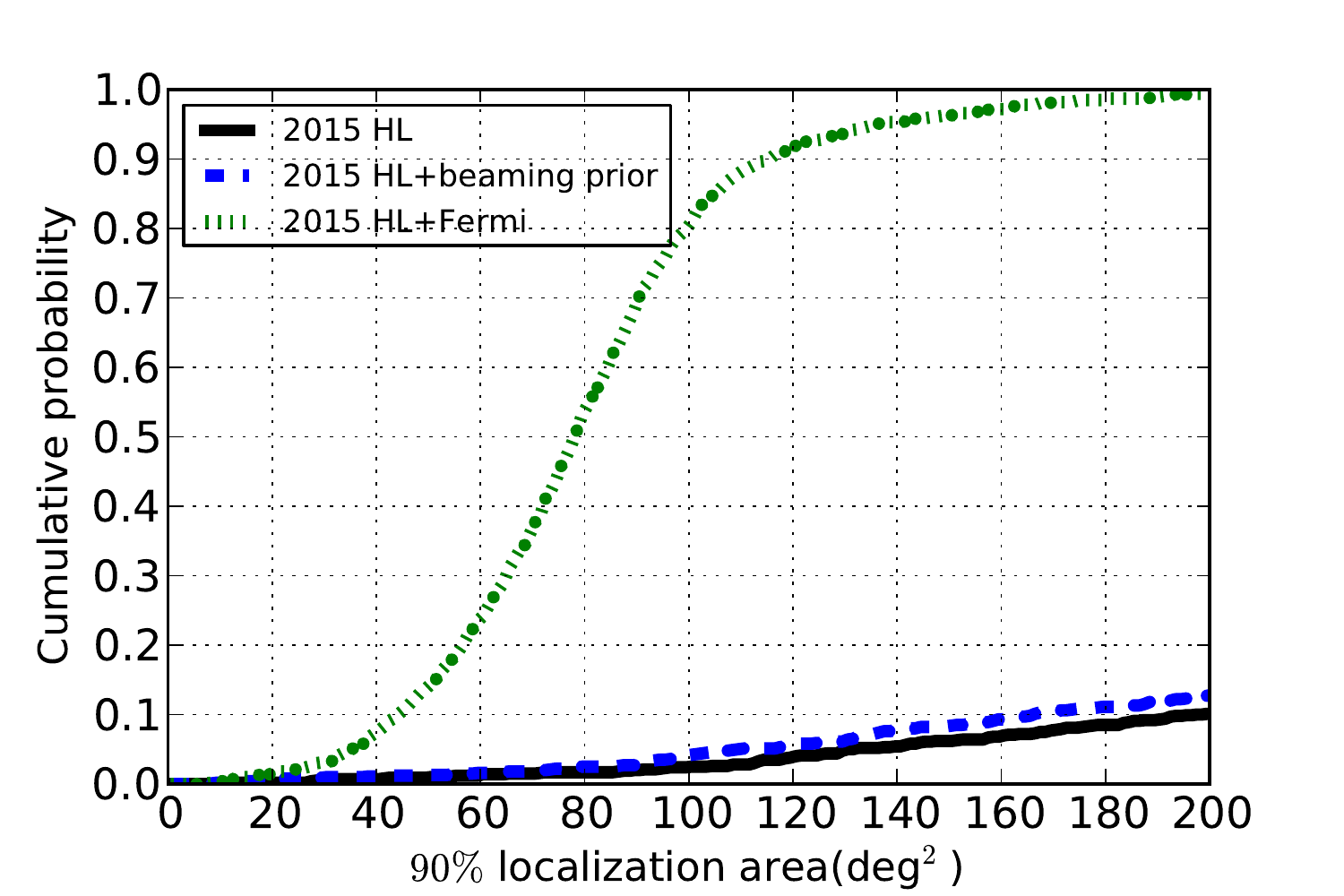}
\caption{\label{fig:fermi}
Cumulative distribution of sky localization for joint GW and Fermi GBM
detections of short GRBs. This shows significant improvement, as compared with
GW detection alone (Table~\ref{table:beaming}). 
}
\end{center}
\end{figure}

In Fig.~\ref{fig:hlv_area} we show the
cumulative distribution of $50\%$ localization areas for the first two
years of advanced LIGO and Virgo network operation. Although
the localization is significantly improved with the addition of Virgo
(c.f. Fig.~\ref{fig:hl_area}),
these represent a minority ($\sim25$\%), since most events detected by this network have a Virgo
SNR below the detection threshold ($\rho_{\rm V}<4$) due to Virgo's
lower sensitivity and mis-aligned antenna
power pattern~\citep{2014arXiv1409.0522C}.
Unless the detection pipeline produces information for these sub-threshold
events, the low-latency localization will be based on information from only two
detectors (i.e., H and L),
 followed by better localization resulting from the full
parameter estimation pipeline analyzing the sub-threshold
waveforms~\citep{2014ApJ...795..105S}; we note that there is a potential remedy to this discussed in Section IX of Singer and Price (2015).
For sources detected with only 2 detectors, 
the localization of events with the 2016 network is potentially worse than those
detected in 2015~\citep{2014ApJ...795..105S}.
This is because the
effective bandwidth of the 2016 noise curve may be narrower than the 2015 case,
leading to inferior localization for
sources with the \emph{same measured}\/ SNRs (e.g., SNR=12). 
Of course, the 2016 network would detect sources from the 2015 network at higher
SNR, and therefore would do a superior job 
of localizing identical sources. Occasionally the event will trigger only H+V or
L+V. In these rare situations the localization will generally be worse than H+L, not only
because Virgo operates at lower sensitivity, but also because the Virgo SNR
distribution skews to lower values when coupled with an H or L detection~\citep{2014arXiv1409.0522C}.


We explore the dependence of localization on SNR, confirming that high SNR
  events lead to better localization. Since there is a universal distribution of
  SNR, we can use this to estimate the benefit of waiting for loud
  events. Fig.~\ref{fig:arearatio} shows that waiting for the loudest out of the
  first 4 events shrinks the median localization area by a factor of almost 2.

Fig.~\ref{fig:fermi} shows that combining an EM prior on the sky
  location of a source with GW localization greatly improves the resulting
  localization. This is because the shapes of the two localization regions are
  ver different and completely uncorrelated. Fermi GBM detection of a short GRB
  would then greatly enhance the localization resulting from a GW network. One
  might have thought that such a detection would also place useful prior 
  constraints on the beaming angle (i.e., gamma-ray detection implies that the
  binary is face-on). However, Table~\ref{table:beaming} and Fig.~\ref{fig:fermi} demonstrate that the
  beaming prior makes only a marginal improvement in localization because of the
  degeneracy between face-on and tail-on binaries.

We have developed a simple, fast, and accurate algorithm for the localization of
GW sources. This algorithm, when coupled with a fast detection pipeline, can play
an important role in the development of multi-messenger astronomy.

\medskip

\begin{acknowledgments}
We acknowledge valuable discussions with Leo Singer, Larry Price, Ben Farr, and Tyson Bailey Littenberg. 
The authors were supported by NSF CAREER grant
PHY-1151836. They were also supported in part by the Kavli Institute for
Cosmological Physics at the University of Chicago through NSF grant PHY-1125897
and an endowment from the Kavli Foundation.
In addition, DEH acknowledges the hospitality of the Aspen Center for
Physics, which is supported by NSF grant PHYS-1066293.
\end{acknowledgments}

\bibliography{ref_loc}

\end{document}